# How transcranial direct current stimulation facilitates post-stroke rehabilitation

Feng Qi[1,2,3], Xin Hong[4], Zhong-Kang Lu[5], Irvin Teh[2], Wei-Peng Teo[6,7], Kai-Keng Ang[5,8], Kok-Soon Phua[5]; Cun-Tai Guan[8]; Effie Chew[9*] and Fatima Nasrallah[2,4,10*]

Current treatments for functional loss of the upper extremity post-stroke remain limited in efficacy, particularly for those with moderate to severe impairment. The non-invasive treatment of transcranial direct current stimulation (tDCS) is a promising intervention technique being tested widely in clinical trials, however, whether tDCS benefits stroke rehabilitation and how tDCS modulates synaptic plasticity are largely unknown. In this study (www.clinicaltrials.gov, NTC01897025), nineteen chronic stroke patients with moderate-to-severe upper limb disability (Fugl-Meyer assessment = 33.9 ± 8.1, out of 66) were recruited for a two-week treatment of combined bi-hemispherical tDCS (anodal electrode placed over affected M1, while cathode over the unaffected M1) and motor imagery based robotic arm training for stroke rehabilitation. Each patient participated in 10 successive rehabilitation sessions, including 20 minutes real/sham tDCS stimulation, followed by 60 minutes of robot assisted arm movement therapy. The primary outcome measure was the upper extremity component of Fugl-Meyer assessment, while the secondary outcome measures included resting motor threshold (RMT) of stroke affected M1 motor cortex, the whole brain longitudinal activity map, and Granger causality graph derived from task-related functional MRI. The RMT measurements showed tDCS evoked higher excitability (lower RMT) in the motor cortex ($\Delta RMT^{tDCS} - \Delta RMT^{sham} = -11.6\%$, $P = 0.04$) that enhanced the descending conduction from the lesioned primary motor cortex to the target hand muscle. Granger causality analysis further revealed the enhanced brain circuitry rewiring from lesioned cerebellum to premotor ($\Delta F_{tDCS}^{Cer\_L->PreM\_R} = +0.0112 \pm 0.0128$, $P = 0.0304$), and from lesioned premotor to primary motor cortex ($\Delta F_{tDCS}^{PreM\_R->PriM\_R} = +0.0077 \pm 0.0100$, $P = 0.0497$) in the tDCS group only owing to the newly formed connections close to the anodal electrode. Rebuilding of these critical pathways was clear via the increase of event related desynchronisation (Laterality$^{tDCS}$ = 0.050, Laterality$^{Sham}$ = -0.063, $P = 0.016$) and white matter integrity in the same lesioned region. Furthermore, only the tDCS group demonstrated a positive recovery trend in the penumbra regions by the longitudinal functional MRI analysis. To interpret tDCS mechanism, we introduce a polarized GABA theory, where GABA$_A$ receptor activity depends on the orientation of dipolar molecule GABA that can be manipulated by tDCS field. Results suggest that tDCS intervention lowers motor excitability via re-orienting GABA, leading to reorganization of the lesioned cortical network, and the motor descending pathway, finally the recovery of motor function.

[1]Oxford Centre for Functional MRI of the Brain, Wellcome Centre for Integrative Neuroimaging, Nuffield Department of Clinical Neurosciences, University of Oxford, UK
[2]Clinical Imaging Research Center, Agency for Science Technology and Research & National University of Singapore, Singapore
[3]Department of Diagnostic Radiology, Yong Loo Lin School of Medicine, National University of Singapore, Singapore
[4]Singapore Bioimaging Consortium, Agency for Science Technology and Research, Singapore
[5]Institute for Infocomm Research, Agency for Science Technology and Research, Singapore
[6]Institute for Physical Activity and Nutrition, Deakin University, Australia
[7]Physical Education and Sports Science Academic Group, National Institute of Education, Nanyang Technological University, Singapore
[8]School of Computer Science and Engineering, Nanyang Technological University, Singapore
[9]Division of Neurology, National University Hospital System, Singapore
[10]The Queensland Brain Institute, The University of Queensland, Australia

Correspondence to: Fatima Nasrallah and Effie Chew
The Queensland Brain Institute, University of Queensland, Australia
Email: f.nasrallah@uq.edu.au
Division of Neurology, National University Hospital System, Singapore
Email: Effie_chew@nuhs.edu.sg

# Introduction

Stroke is one of the main causes of acquired adult disability (Langhorne *et al.*, 2011) resulted from brain infarction or ischemia in territory supplied by cerebral arteries. It leads to a functional disruption in brain circuitry preventing flow of information across brain regions, meanwhile, it also triggers a cascade of synaptic plasticity events from molecular to systems behavior (Murphy and Corbett, 2009) to recover from the injury. Rehabilitation therapy such as robot-assisted MI-BCI training is effective to enhance motor functionality, with activation of neural pathways similar to that of motor execution. And our previous studies (Ang *et al.*, 2008, 2011) have demonstrated feasibility and efficacy of motor imagery-based brain computer interface for post-stroke motor impairment, in which MI is detected by surface EEG and translated to execution of the target movement with the aid of an arm robot (MIT-Manus, Interactive Motion Technologies). In this study, we aim to test whether the concurrent therapies of MI-BCI and tDCS could improve motor recovery of post-stroke patients.

There is still no clear conclusion whether tDCS could benefit cognition enhancement or facilitate post-stroke recovery. Some groups demonstrated the potential of tDCS intervention in treating schizophrenia (Orlov *et al.*, 2017), central alexia (Woodhead *et al.*, 2018) and stroke recovery (Stagg *et al.*, 2012), while others reported no good evidence that tDCS is useful (Agarwal *et al.*, 2013; Tremblay *et al.*, 2014; Cooney *et al.*, 2015). Indeed, all these uncertainties stem from the question of how tDCS interact with neural system. One widely accepted interpretation of tDCS mechanism is 'intracranial current flow can alter neuronal activity' (Cavaleiro *et al.*, 2006; Nitsche *et al.*, 2008), however, how current flow makes such alteration is not explained. In this study, by comparing the results of real versus sham tDCS treatments with multi-modality measurements, we aim to figure out how neural circuitry rewire under tDCS intervention.

For assessment of motor impairment and rehabilitation performance, clinical trials have widely used FMA score (Lee *et al.*, 2001; Gladstone *et al.*, 2002a). However, the questionnaire-based impairment index is limited in the reliability and validity owing to the subjectivity of participant perception. Nevertheless, as modern neuroimaging techniques advance, therapy assessments tend to involve more objective and precise measures, such as electromyography for muscle cell activity (Thaut *et al.*, 1997), and MRI/EEG for neuronal responses and connectivity (Wu *et al.*, 2006, 2015; Grefkes and Fink, 2011). These techniques make it possible to quantitatively characterize the difference of brain states between real and sham tDCS interventions in term of image derived phenotypes (Smith *et al.*, 2015; Elliott *et al.*, 2018) such as white matter integrity and functional connectivity across brain areas. In this study, we aim to introduce more effective imaging biomarkers for evaluating tDCS intervention and predicting its prognosis.

## Material and methods

### Study design

We conducted a randomized, double-blinded study of combined real-tDCS and MI-BCI vs sham-tDCS and MI-BCI in chronic stroke patients which spanned over a full duration of 4 weeks (Fig.1). Each patient participated in ten successive rehabilitation sessions, including 20 minutes real/sham tDCS stimulation, followed by 8 minutes online MI accuracy measurement, and then 60 minutes of MI-BCI therapy. Measures of functional assessment were also collected at multiple time points mainly at week 0, week 2 and week 4. Patients were scheduled for two separate MRI scans at pre-training (week 0) and post-training (week 4). Since spontaneous recovery typically plateaus within 3 to 6 months (Langhorne *et al.*, 2011), the observed motor improvements in subjects (9+ months post-stroke in Table 1) is expected to be attributed to applied treatments.

### Study subjects

Ethics approval was obtained from the Institution's Domain Specific Review Board, National Healthcare Group, Singapore. This trial was registered in www.ClinicalTrials.gov (NCT01287975). Informed consent was obtained prior to study enrollment. Forty-two subjects were assessed for eligibility, seven declined to participate, eleven failed to BCI screening, five were excluded due to a history of seizures, major depression, and metal implants that were not suiTable for MRI, and one was excluded because lesion extended to motor cortex. In total, eighteen subjects (age = 54.5 ± 10.8, females 54.6 ± 12.2, males 54.5 ± 10.7) were eventually recruited into the study and completed the training. Eleven age matched healthy volunteers (age = 56.7 ± 4.5) were also recruited.

### tDCS intervention

A randomized stratification approach, with a computer-generated random sequence, was used to group subjects according to their FMA scores and then randomize them according to a block randomization to receive either real (n = 9, $FMA_{wk0}$ = 36 ± 8) or sham (n = 9, $FMA_{wk0}$ = 32 ± 8) tDCS, with no group difference in age, sex, stroke type, stroke nature, affected limb, post-stroke time at week 0 (Table 1).

Bilateral tDCS of 1 mA for 20 minutes was delivered to the motor cortex with two surface sponge electrodes with the anode electrode placed on the ipsilesional M1 position of the brain while the cathode on the contralesional part. This is to decrease cortical excitability in unaffected motor cortex and to increase it in the affected motor cortex (Lindenberg *et al.*, 2010). Constant current was supplied by a battery-operated current stimulator (neuroConn DC Stimulator GmbH). The sham group, on the other hand, had the same setup, but was administered at 1 mA tDCS for 30 seconds instead.

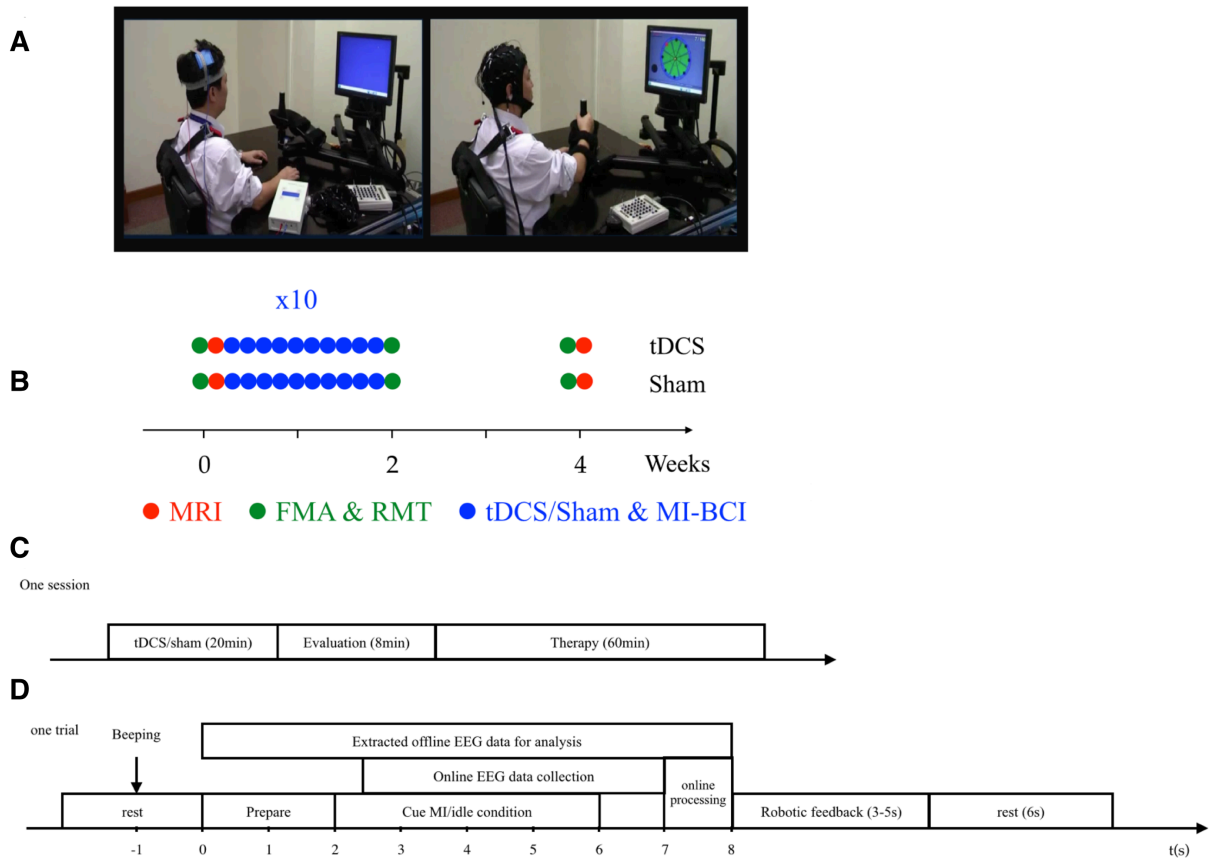

**Figure 1 Study design**

## Table 1 Demographics and clinical data

| Subject ID | Age (years) | Gender | Post stroke time (months) | Stroke type | Affected arm | FMA score | | | RMT | | |
|---|---|---|---|---|---|---|---|---|---|---|---|
| | | | | | | Pre (wk0) | Post1 (wk2) | Post2 (wk4) | Pre (wk0) | Post1 (wk2) | Post2 (wk4) |
| **tDCS Group** | | | | | | | | | | | |
| N001 | 29 | M | 12 | Infarct | L | 51 | 50 | 51 | 76 | 64 | 68 |
| N005 | 54 | M | 28 | Infarct | R | 29 | 34 | 42 | 86 | 78 | 57 |
| N006 | 38 | F | 29 | Hemorrhage | L | 38 | 41 | 42 | — | — | — |
| N010 | 60 | F | 51 | Hemorrhage | L | 26 | 22 | 31 | 80 | 74 | 68 |
| N015 | 48 | F | 49 | Hemorrhage | R | 39 | 42 | 46 | — | — | — |
| N025 | 59 | M | 13 | Infarct | R | 31 | 28 | 31 | 79 | — | 72 |
| N027 | 65 | M | 27 | Infarct | R | 41 | 45 | 48 | — | — | — |
| N029 | 57 | F | 10 | Hemorrhage | R | 40 | 40 | 44 | — | — | — |
| N037 | 64 | M | 86 | Infarct | L | 28 | 29 | 28 | — | — | 71 |
| **Sham Group** | | | | | | | | | | | |
| N007 | 51 | M | 44 | Infarct | L | 33 | 42 | 45 | 80 | 93 | 90 |
| N009 | 39 | M | 25 | Infarct | R | 36 | 42 | 39 | — | — | — |
| N011 | 59 | M | 52 | Hemorrhage | L | 41 | 46 | 57 | 91 | 89 | 84 |
| N018 | 70 | F | 19 | Infarct | L | 23 | 25 | 26 | — | — | — |
| N019 | 59 | M | 44 | Infarct | L | 29 | 24 | 28 | — | — | — |
| N021 | 58 | M | 29 | Infarct | R | 28 | 32 | 37 | — | — | — |
| N030 | 58 | M | 25 | Hemorrhage | L | 20 | 22 | 24 | — | 90 | 90 |
| N031 | 47 | M | 10 | Infarct | R | — | 40 | 40 | — | — | — |
| N032 | 66 | M | 52 | Infarct | L | 43 | 45 | 46 | 60 | 75 | 65 |
| **Patient mean (SD)** | 55 (11) | — | 34 (20) | — | — | 34 (8) | 36 (9) | 39 (9) | 79 (10) | 80 (11) | 74 (12) |
| **Patient range** | 29-70 | — | 10-86 | — | — | 20-51 | 22-50 | 24-57 | 60-91 | 64-93 | 57-90 |
| **tDCS mean (SD)** | 53 (12) | — | 34 (25) | — | — | 36 (8) | 37 (9) | 40 (8) | 80 (4) | 72 (7) | 67 (6) |
| **Shame mean (SD)** | 56 (9) | — | 33 (15) | — | — | 32 (8) | 35 (10) | 38 (11) | 77 (16) | 87 (8) | 82 (12) |

## MI-BCI training

Following the tDCS session, MI-BCI therapy were performed for 160 trials (~1 hour) on each patient. The MIT-Manus robot (Interactive Motion Technologies) was used to facilitate the upper extremity movement. It is a 2 degree of freedom robotic exoskeleton that is controlled by the motor imagery signal acquired by a 32 channel NeuroScan NuAmps EEG Amplifier (Compumedics USA), of which 27 channels are used, and translated by Filter Bank Common Spatial Pattern (FBCSP) algorithm (Ang *et al.*, 2012). In the evaluation portion (Fig.1C), quantitative online MI-BCI performance was acquired; while, in the therapy portion, quantitative μ-ERD (8-13Hz) was obtained (Ang *et al.*, 2015).

## Primary outcome measure

Clinical measures were performed (Week 0, Week 2, and Week 4) to evaluate the rehabilitation effect. The upper extremity score of the FMA (Gladstone *et al.*, 2002b) which assessed upper extremity motor impairment including balance, coordination and speed of arm, wrist, finger was measured for primary motor functional performance.

## Secondary outcome measure

### RMT assessment

RMT of affected arm was measured by single-pulse transcranial magnetic stimulation (TMS) at the same time points of Week 0, Week 2, and Week 4. RMT is determined as the percentage of the maximum stimulator output required to elicit motor evoked potential (MEP) with 50–100μV peak-to-peak amplitude using Bistim 2002 (Magstim Co., UK). A lower stimulator output to elicit RMT indicates a higher degree of corticospinal excitability (Rossini *et al.*, 2015; Hong *et al.*, 2017).

### MRI assessment

MR scans were performed on a 3T Tim Trio scanner (Siemens, Germany) with a 32 channel head array coil. Subjects lay supine in the scanner with a rubber bulb in each hand and resting in custom hand supports to prevent movement. A T1-weighted anatomical image was acquired with a magnetization prepared rapid gradient-echo (MPRAGE) sequence with the following parameters: TR = 1900 ms, TI = 900 ms, TE = 2.52 ms, FOV=256 * 256, flip angle = 8, number of slices = 176 , voxel size = 1 mm isotropic, and slice thickness = 1 mm. Functional MRI data was acquired during finger griping task using a single shot Echo Planar Imaging (EPI) sequence with using the parameters: TR = 3000 ms, TE = 35 ms, FOV= 182 * 182, voxel size = 3.4 mm isotropic, and number of measurements = 89. The detailed methods and results of T2 weighted MRI and diffusion weighted MRI can be referred to our report (Hong *et al.*, 2017).

Image preprocessing was implemented by a combination of the FMRIB Software Library (Smith, 2002; Jenkinson *et al.*, 2012) and ANTS registration tools (Petoud *et al.*, 2008). This included brain extraction by BET, motion correction by MCFLIRT, spatial smoothing by a Gaussian Kernel of 8.0 mm, and temporal filtering with high pass filter cutoff of 60 s. Permutation tests were performed with threshold-free cluster enhancement and family-wise error rate controlled to a significance level of $P < 0.05$. Seven patients had stroke cores in the left hemisphere, and their functional maps were flipped horizontally so that all 'left hand active' tasks related to lesion on the right motor cortex and left cerebellum for all subjects. First level analysis of the fMRI data was performed with FSL-FEAT (Woolrich *et al.*, 2001).

| Healthy (R) | Pre-train (G) | post-train (B) | Encoded Color | Color Voxel Meaning |
|---|---|---|---|---|
| 0 | 0 | 0 | Black | Task unrelated voxels |
| 1 | 0 | 0 | Red | Permanent damaged voxels |
| 0 | 1 | 0 | Green | Short term compensatory voxels |
| 1 | 1 | 0 | Yellow | Function lose during rehabilitation |
| 0 | 0 | 1 | Blue | Function acquired during rehabilitation |
| 1 | 0 | 1 | Purple | Function recovered during rehabilitation |
| 0 | 1 | 1 | Lightblue | Long term compensatory voxels |
| 1 | 1 | 1 | White | Voxels without stroke influence |

**Table 2 RGB color codes**

**Longitudinal RGB voxel analysis**

The functional z-score maps were calculated for healthy control and patient groups at pre-training (week 0) and post-training (week 4) with FSL-FEAT, 1st level analysis. To facilitate longitudinal studies, different groups were superimposed in RGB style and were color coded into color vector [R G B] for simple representation and delineation (Table 2). The averaged functional map of control subjects was rendered in Red component, patients' pre-training in Green, and post-training in Blue. By comparing the Red (healthy control) and Green (Pre-training) functional maps, the stroke influence could be obtained; while by comparing the Green (pre-training) and Blue (Post-training) functional maps, rehabilitation effects can be concluded. Group functional maps from all three groups were superimposed to determine the differences in the extent of functional activation induced by the task. To facilitate interpretation of the findings, Table 2 was used to present each color-coded information. Quantitative longitudinal analysis of the functional activation was assessed with voxel statistics. For the RGB quantitative analysis, the motor related areas including primary motor cortex (PriM), premotor cortex (PreM), supplementary motor area (SMA), and cerebellum were analyzed statistically by extracting functional data from FSL atlas. Each individual patient's functional RGB image and voxel number of different color were calculated.

**Granger Causality analysis**

Multivariate Granger causality analysis (MVGC) (Seth, 2010; Barnett and Seth, 2014) was conducted using Matlab (Math-works, Natick, US). Functional time series were extracted from specified regions of interest which included both left and right PriM, PreM, SMA, dorsal lateral prefrontal cortex (PFC), superior parietal cortex (SPC), insular cortex (Ins), visual cortex (VC), primary sensory cortex (S1), thalamus (Tha), basal ganglia (BG) and Cerebellum (Cer). Tha, BG, and Cer were divided into sub-functional regions, and the hand related sub-regions were used for presentation. The extracted time series of each ROI were fed to the MVGC engine as input. The ordinary least squares (OLS) was used for solving the vector autoregressive (VAR) model, Akaike information criterion (AIC) was used to select the model order, and nonparametric bootstrap resampling was used on generated VAR data (sample number m = 702) (Barnett and Seth, 2014). Statistical correction for significance employed the false discovery rate (FDR) correction (alpha = 0.01). The final pairwise-conditional Granger causality were computed by MVGC toolbox (Barnett and Seth, 2014). The longitudinal comparison study of the causality matrix among different groups was then used to assess the rehabilitation results. 22-node G-causality graphs were manually drawn by Keynotes (Apple Inc., US), 3-node and 5-node G-causality graphs by Matlab.

**Statistical analyses**

To evaluate the tDCS effect, the change of measures (FMA, RMT, MRI functional map and Granger causality) before and after rehabilitation were calculated for each patient. Then, we quantitatively characterized these changes between the real and sham tDCS in term of effect size (Cohen's d), and tested the significance using two sample unequal variance t-test by Numbers (Apple Inc., US), where alpha = 0.05 was taken as default for statistical significance.

## Data availability

The data that support the findings of this study are available on request from the corresponding author.

## Results

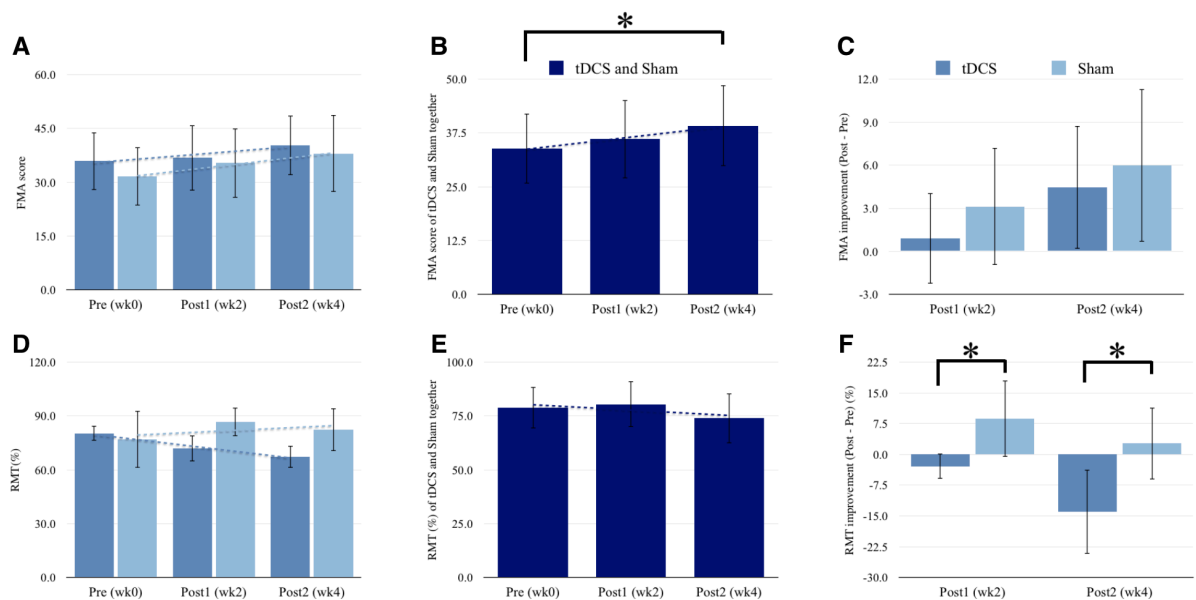

**Figure 2 Statistics of FMA (A-C) and RMT (D-F)**

### FMA score

Fig.2 A-B demonstrate both tDCS and sham group experienced positive FMA trends ($slope_{tDCS}$ = +2.22, $R^2_{tDCS}$ = 0.89; $slope_{sham}$ = +3.19, $R^2_{sham}$ = 0.99) due to the MI-BCI rehabilitation, and FMA of whole stroke participants in week 4 gained significant improvement compared to pre-training ($\Delta FMA_{wk4}$ = +5.3, $P_{wk4}$ = 0.04). FMA improvements immediate after rehabilitation and 2 weeks follow-up were observed ($\Delta FMA^{tDCS}_{wk2}$ = 0.9 ± 3.1, $\Delta FMA^{sham}_{wk2}$ = 3.1 ± 4.1,

$\Delta FMA^{tDCS}_{wk4} = 4.4 \pm 4.3$, $\Delta FMA^{sham}_{wk4} = 6.0 \pm 5.3$) in Fig.2C, however, there was no significant difference of FMA improvements between sham and tDCS intervention ($P_{wk2} = 0.23$, $P_{wk4} = 0.50$).

**RMT score**

Fig.2 D demonstrates tDCS treatment induced a negative trend of RMT value (slope$_{tDCS}$ = -6.5, $R^2_{tDCS}$ = 0.98), while sham treatment had positive trend (slope$_{sham}$ = +2.63, $R^2_{sham}$ = 0.29). Here, we note that lower RMT value implies higher excitability in motor cortex (Khedr EM *et al.*, 2005; Ward *et al.*, 2006). The RMT of whole stroke participants (Fig.2 E) also shows a declining trend due to MI-BCI rehabilitation (slope$_{whole}$ = -2.48, $R^2_{whole}$ = 0.53). Particularly, as compared with sham, tDCS group illustrates significant treatment effect in term of RMT drop (Fig.2F) in both week 2 and week 4 (Cohen's $d_{wk2}$ = -1.73, $P_{wk2}$ = 0.04; Cohen's $d_{wk4}$ = -1.33, $P_{wk4}$ = 0.03).

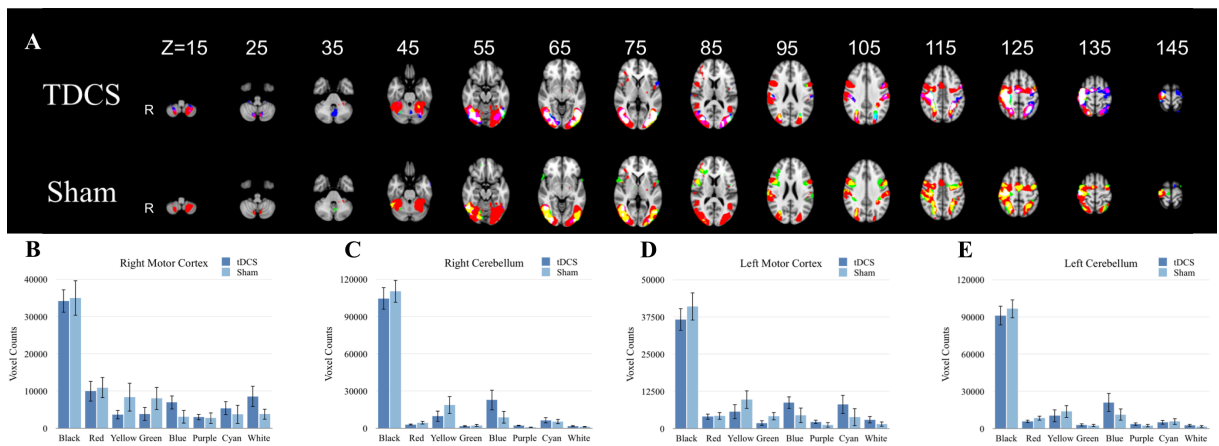

**Figure 3 Longitudinal fMRI analysis in RGB style**

**Longitudinal RGB analysis**

A RGB longitudinal representation of the functional activity during left hand active task is presented in Fig.3 A, with the color encoded information displayed in Table 2, and statistic bar chart in Fig.3B-E. The group RGB functional voxel statistics were calculated from individual fMRI maps within sham or TDCS groups. tDCS functional maps have more blue (+3829; +9779), purple (+290; +1259) and white (+4779; +833) voxels, while less red (-912; -2567) and yellow (-4673; -3462) voxels, in form of voxel difference in (right motor; left cerebellum) (Supplementary Table 1). And the enlarged longitudinal RGB maps in motor area are shown in Supplementary Fig. 1.

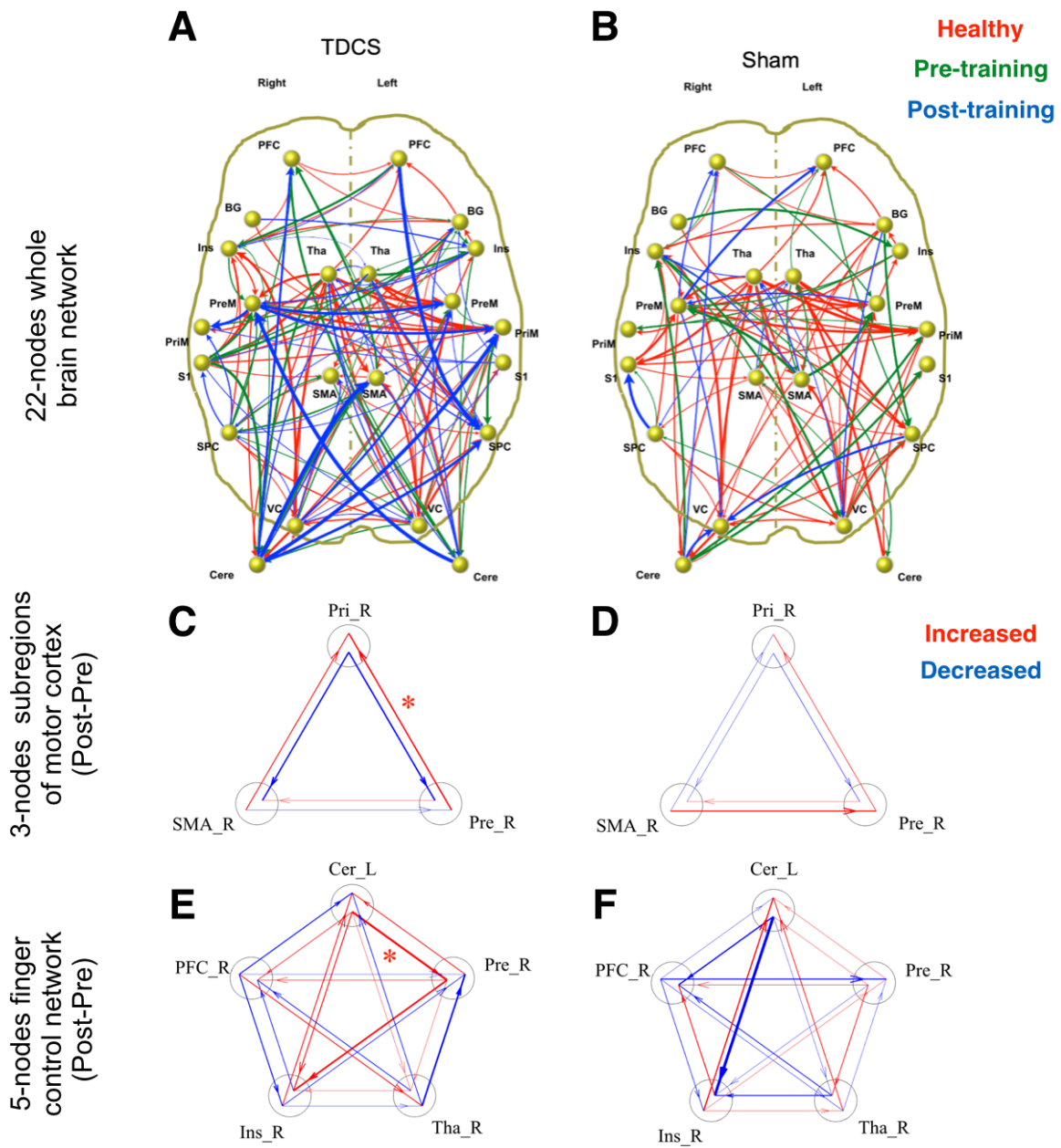

**Figure 4 Granger causality graphs under left hand active task.**

A representation of the cortical network extracted from the MVGC analysis (Seth, 2010; Barnett and Seth, 2014) is shown in Fig. 4A-B and the Granger causality matrix is shown in Supplementary Fig. 2. Significant visual-motor related causality connections (i.e. information flow under gripping task) were presented within the healthy control group (red arrows, $F^{control}$ = 0.4490 ± 0.1120) that were lost in both the sham ($F_{wk0}^{sham}$ = 0.1683 ± 0.0310) and tDCS ($F_{wk0}^{tDCS}$ = 0.2571 ± 0.0343) patient groups after stroke in pre-training condition (green arrows).

After rehabilitation (blue arrows), tDCS revealed an increase of information flow ($\Delta F_{wk4-wk0}^{tDCS}$ = +0.0563 ± 0.0862), while sham experienced a decrease of information flow ($\Delta F_{wk4-wk0}^{sham}$ = -0.0669 ± 0.0376), and thus the tDCS intervention effect could be quantitatively characterized as compared to sham (Cohen's $d^{tDCS}$ = +1.3357, P = 0.0035). In particular, significant afferent and efferent causality connections were observed in the affected motor cortex and cerebellum compared to the sham group. For instance, in post-rehabilitation week 4, the total significant efferent causalities from the right premotor cortex (PreM_R) were $F_{tDCS\_wk4}^{fromPreM\_R}$ = 0.0396 ± 0.0281 vs $F_{Sham\_wk4}^{fromPreM\_R}$ = 0.0202 ± 0.0103, afferent causalities to PreM_R were $F_{tDCS\_wk4}^{toPreM\_R}$ = 0.0259 ± 0.0165 vs $F_{Sham\_wk4}^{toPreM\_R}$ = 0.0076 ± 0.0049; while the efferent causalities from the left cerebellum (Cere_L) were $F_{tDCS\_wk4}^{fromCere\_L}$ = 0.0240 ± 0.0134 vs $F_{Sham\_wk4}^{fromCere\_L}$ = 0.0016 ± 0.0011, the afferent causalities to the Cere_L were $F_{tDCS\_wk4}^{toCere\_L}$ = 0.0141 ± 0.0081 vs $F_{Sham\_wk4}^{toCere\_L}$ = 0 (no significant connection existed), as is shown in Supplementary Table 2. In addition, our analysis of the Granger causality connectivity following MI-BCI/tDCS training, revealed a remarkably stronger information flow from the right pre-motor (PreM_R) to left pre-motor (PreM_L) in the Real tDCS group ($F_{tDCS\_wk4}^{PreM\_R->PreM\_L}$ = 0.0131 ± 0.0113, P = 0.0084) than in the Sham group ($F_{sham\_wk4}^{PreM\_R->PreM\_L}$ = 0.0083 ± 0.072, P = 0.0224).

To further investigate the tDCS intervention effect in detail, Fig. 4 C-F and Supplementary Table 3 demonstrate the causality connection changes among 3-node motor subregions network (supplementary motor area, pre-motor and primary motor cortex) and 5-node figure control network (supplementary motor area, thalamus, insular, pre-motor, and primary motor cortex). Among the fifty-two connections, only two significant alterations were found and both in the tDCS group ($\Delta F_{tDCS}^{Cer\_L->PreM\_R}$ = +0.0112 ± 0.0128, P = 0.0304; $\Delta F_{tDCS}^{PreM\_R->PriM\_R}$ = +0.0077 ± 0.0100, P = 0.0497)

**Discussion**

This study tested the effect of combined tDCS and MI based robotic arm training for stroke rehabilitation, gained insight into the neurophysiological mechanism by comparing cortical excitability changes following sham-tDCS/BCI versus real-tDCS/BCI, and proposed a longitudinal RGB MRI analysis method for assessing tDCS intervention efficacy and predicting its prognosis.

The upper extremity FMA score is one of the most widely used clinical measure to assess post-stroke rehabilitation effect of MI based robotic arm training (Klamroth-Marganska *et al.*, 2014) and tDCS intervention(Allman *et al.*, 2016). Our results in Fig.1 A-C illustrate that both tDCS and sham improved participants' motor functionality immediate after rehabilitation and 2-week follow-up ($\Delta FMA^{tDCS}_{wk2}$ = 0.9 ± 3.1, $\Delta FMA^{sham}_{wk2}$ = 3.1 ± 4.1, $\Delta FMA^{tDCS}_{wk4}$ = 4.4 ± 4.3, $\Delta FMA^{sham}_{wk4}$ = 6.0 ± 5.3). In particularly, the week 4 FMA of cohort participants showed significant improvement ($\Delta FMA_{wk4}$ = +5.3 ± 4.7, $P_{wk4}$ = 0.04), as compared with pre-training. All these evidenced that the rehabilitation of MI based robotic arm could benefit muscle movement of post-stroke participants, especially for those with severe impairment. In addition, such beneficiary effect could be maintained or even be more remarkable in the follow-up weeks.

From Fig.2C, we also witness that FMA score could not differentiate between sham and tDCS intervention ($P_{wk2}$ = 0.23, $P_{wk4}$ = 0.50). This is due to the subjectivity and inaccuracy of FMA's questionnaire based test (Van Delden *et al.*, 2013), that's why many objective measures such as electrophysiology based RMT and reaction time measurement, and MRI based connectivity and diffusivity measurements have been adopted clinically (Murase *et al.*, 2004; Saleh *et al.*, 2004; Hummel *et al.*, 2009; Adhikari *et al.*, 2017).

Fig. 2D demonstrates that the tDCS intervention significantly decreased the RMT at both immediate post-training (Cohen's $d_{wk2}$ = -1.73, $P_{wk2}$ = 0.04) and 2-week follow-up (Cohen's $d_{wk4}$ = -1.33, $P_{wk4}$ = 0.03) as compared to the sham group. This is consistent with our previous results from diffusion tensor imaging in the same cohort of patients where fractional anisotropy (FA) was significantly increased in the ipsi-lesional corticospinal tract (cst) (Hong *et al.*, 2017). Integrating the reduced RMT values, as an indicator of increased conductivity (Pfurtscheller and Lopes, 1999), and the increased FA in the tDCS-only group provides direct evidence of improved neural pathways from lesioned primary motor cortex to the abductor pollicis brevis (APB) muscle of the affected hand via cst and descending path in spinal cord.

Longitudinal MRI is presented in Fig. 3 in RGB style, whose color codes are defined in Table 2. The results illustrate that tDCS group has more blue (+3829; +9779), purple (+290; +1259) and white (+4779; +833) voxels in the affected (right motor; left cerebellum) regions, which indicates tDCS facilitates the ability to recruit, repair, and maintain functioning voxels, while sham group has more red (+912; +2567) and yellow (+4673; +3462) voxels that indicates a trend of losing functioning voxels (Supplementary Table 1). The enlarged version of motor cortex in Supplementary Fig. 1 shows that tDCS treatment has a positive recovering potential because its red (damaged functional voxels) are bounded by purple (recovered functional voxels), while sham tends to lose more white (unaffected functioning voxels) surrounded by yellow (function lost during training). Higher excitability of tDCS on the motor cortex is also presented in the RGB fMRI graph, where remarkable higher blue hue (blue, purple, cyan and white) voxels, representing more task related activity occurring in post-training stage. this is consistent with our μ-ERD study (Ang et al., 2014, 2015), whose laterality coefficient analysis also demonstrates that, after rehabilitation, ipsilesional motor cortex had significantly more activated desynchronized neurons in tDCS group (Laterality$^{tDCS}$ = +0.050) than in sham group (Laterality$^{sham}$ = -0.063, P = 0.016). The increase in the number of blue hue voxels and desynchronized neurons suggests a rise of the brain activity in affected motor cortex and enhanced capacity of synaptic reorganization in that region according to Hebbian learning rule (Wong and Ghosh, 2002). So, the RGB style longitudinal fMRI analysis revealed that tDCS intervention added more benefits in treatment effects and future prognosis than sham.

Information flow under hand gripping task could be derived from MVGC analysis by feeding the time series data of functional MRI (Seth, 2010; Barnett and Seth, 2014). Fig. 4A-B show that the healthy control group (red arrows, $F^{control}$ = 0.4490 ± 0.1120) possessed stronger Granger causality connections than either sham ($F_{wk0}^{sham}$ = 0.1683 ± 0.0310) or tDCS ($F_{wk0}^{tDCS}$ = 0.2571 ± 0.0343) patient groups (green arrows), which suggests these subcortical strokes had severely disabled information communication among brain regions. After rehabilitation (blue arrows), we compared tDCS and sham in term of change of Granger causality (or information flow under gripping task) ($\Delta F_{wk4-wk0}^{tDCS}$ = +0.0563 ± 0.0862 versus $\Delta F_{wk4-wk0}^{sham}$ = -0.0669 ± 0.0376), and witnessed a significant tDCS intervention effect as characterized by Cohen's $d^{tDCS}$ = +1.3357, P = 0.0035. In particular, the improvement of causality connections in the affected motor cortex and cerebellum were observed as compared to the sham group (Supplementary Table 2). Stronger Granger causalities in tDCS group indicate its positive effects in brain circuitry rebuilding in critical motor areas during post-stroke training. Especially, the stronger inter-hemispherical Granger causality from PreM_R to PreM_L in the Real tDCS than sham group ($F_{tDCS\_wk4}^{PreM\_R->PreM\_L}$ = 0.0131 ± 0.0113, $F_{sham\_wk4}^{PreM\_R->PreM\_L}$ = 0.0083 ± 0.072) is also confirmed by the increased FA of corpus collosum (cc) connecting the bihemispherical premotor cortices reported from the same cohort in our recent study (Hong et al., 2017). The enhanced information flow and white matter tract integrity demonstrate the beneficial effect of tDCS in rebuilding of connectivity and communication between two hemispheres (Carter *et al.*, 2010).

Interactions among supplementary motor area, pre-motor and primary motor cortex, of the right hemisphere are key in the planning and execution of left active finger pinching task (Nachev *et al.*, 2008; Ebbesen and Brecht, 2017). We compared the causality change (post-training – pre-training) between tDCS and sham groups, and found that the only significant change was the causality increase ($\Delta F_{tDCS}^{PreM\_R->PriM\_R} = +0.0077 \pm 0.0100$, P = 0.0497) in tDCS from the right PreM to right PriM (Fig. 4C-D and Supplementary Table 3). The interactions among left cerebellum, right cerebral cortex and thalamus are key in accurate control of finger action (Ramnani, 2006). From the 5 node causality change graph (Fig. 4E-F and Supplementary Table 3), we found that the only significant causality change ($\Delta F_{tDCS}^{Cer\_L->PreM\_R} = +0.0112 \pm 0.0128$, P = 0.0304) in tDCS was from left cerebellum to right pre-motor cortex, which sends error information of finger motion to the cortex to fine control the pinching action (Bostan and Strick, 2018). So the increased excitability in motor cortex induced by tDCS facilitates the circuit rebuilding among cortical regions and cerebellum.

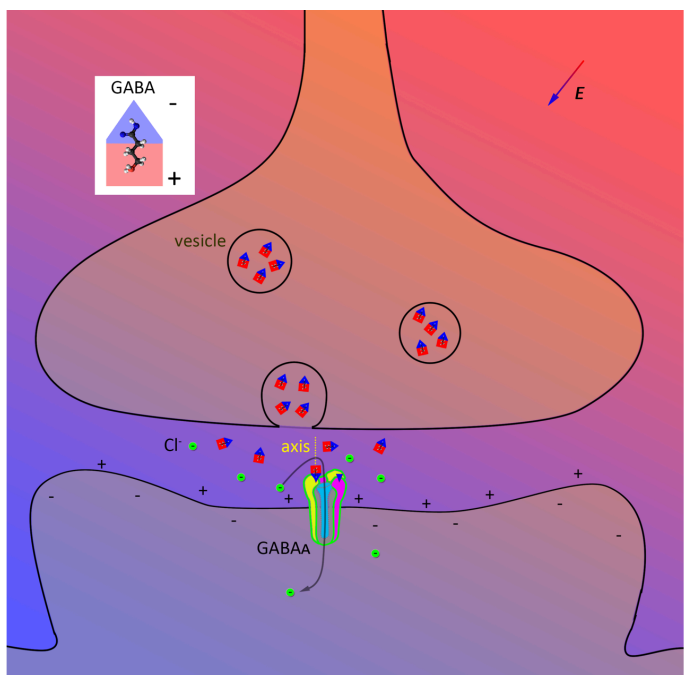

**Figure 5. Polarized GABA theory for tDCS effect.**

For tDCS mechanism, commonly accepted interpretation is 'intracranial current flow can alter neuronal activity' (Cavaleiro *et al.*, 2006; Nitsche *et al.*, 2008), but how current flow makes such alteration is not explained. Instead, we proposed a polarized GABA hypothesis as illustrated in Fig. 5. The logic is that synaptic plasticity requires neurons to be activated that in turn activate N-methyl-D-aspartate receptors (NMDAR) leading to influx of $Ca^{2+}$ into dendrites to activate the plasticity process via kinases such as $Ca^{2+}$/calmodulin-dependent protein kinase II (CAMKII) (Roelfsema and Holtmaat, 2018). To enhance neuronal activity, the external exerted electric field can make benefit. Since GABA is a small electric dipolar molecule with one negative carboxyl end and one positive amino end, external electric field can reorient GABA along the field direction. If the polarized GABA aligned with pocket axis of GABA docking site in $GABA_A$ receptors (i.e. the 'aromatic box' in $\beta^+/\alpha^-$ interface) (Zhu *et al.*, 2018; Laverty *et al.*, 2019; Masiulis *et al.*, 2019), external electric field can benefit GABA inhibition function via allowing $Cl^-$ inflow, the pyramidal neuron will be less excitable. If the external filed makes GABA opposite or misaligned (for most of the time) to receptor pocket axis, pyramidal neurons will increase the excitability. The polarized GABA theory elucidates the intervention mechanism of tDCS, and suggests a novel approach to manipulate neuronal electrophysiology by external electric field.

In sum, there are four components of necessity for stroke recovery under tDCS intervention. First tDCS elevated motor cortical excitability via misaligning GABA with its receptors by external tDCS field. The elevated excitability is evidenced by the increase of event-related desynchronization (Ang *et al.*, 2015), and by the more blue, purple and white voxels in tDCS than sham group (Fig. 3). Then, the tDCS-elevated excitability facilitated neural circuitry rebuilding through strengthening the associated connectivity from cerebellum to premotor, and from premotor to primary motor cortex evidenced by changes in the causality strength between these regions (Fig. 4). Finally, neural circuitry rebuilding through strengthening the neural descending pathway from primary motor cortex to target muscle evidenced by decreased RMT (Fig. 1) and increased corticospinal tract integrity by diffusion MRI measurements (Hong *et al.*, 2017).

Together, this study has demonstrated for the first time the synergistic effect of tDCS and MI-BCI on improvement of upper limb function evidenced by measurements of EEG, TMS/MEG and MRI. We also proposed a longitudinal RGB fMRI analysis which not only confirmed that, with tDCS treatment, post-stroke patient could better maintain, recover and develop functioning voxels in ipsi-lesional motor cortex and cerebellum, but also predicted positive prognosis by tDCS treatment. With the proposed polarized GABA theory, we illuminated how tDCS intervened in the neural plasticity, and how neural circuitry reorganized from affected cerebellum to motor cortex finally to the target muscle.

## Funding

This trial (Clinical trial registration No. NCT01897025) was supported by the Science and Engineering Research Council of Agency for Science, Technology and Research, and the National Medical Research Council, Singapore (Grant No. NIG09may022), and by Motor Accident Insurance Commission, The Queensland Government, Australia (Grant No. 2014000857)

## Supplementary material

Supplementary material is available at *Brain* Online.

**Figure Legends**

**Table 1. Demographics and clinical data**. Note: M = male; F = female. L = left, R = Right. Subject #35 was excluded because lesion extended to motor cortex. Subject #31 was not available for FMA and RMT measurements in week 0. Subject #25 did not get the second RMT measurement due to technical difficulty. All other '−' labels in the RMT reading denote undetectable motor evoked potential.

**Table 2. RGB color codes**. R component represents averaged functional activity under task in healthy state, while G component represents the pre-training brain activity (after stroke), and B component represents the post-training brain activity. With the superimposed RGB components as input, each voxel can be rendered into 8 colors with diagnostic meanings.

**Figure 1. Study design.** (**A**) patient undergoes tDCS intervention and MI-BCI with robotic arm feedback rehabilitation. The same setup for sham. (**B**) The rehabilitation lasted from week 0 to week 2, both sham and real tDCS groups conducted 10 sessions of sham/tDCS + MI-BCI therapy. Screening assessments were not drawn here. (**C**) Each rehabilitation session contains 20 minutes tDCS/sham stimulation, followed by 8 minutes online MI accuracy evaluation, and 60 minutes 160-trial MI-BCI with robotic feedback rehabilitation. (**D**) Timing of the motor imagery of the stroke-affected hand using online MI-BCI with robotic feedback for the rehabilitation session. The quantitative μ-ERD was calculated by offline EEG analysis.

**Figure 2. Statistics of FMA (A-C) and RMT (D-F)** across pre-training (week 0), immediate post-training (week 2) and 2-week follow-up (week 4). Bar charts are presented in term of (mean ± standard deviation), and * indicates P < 0.05. Note: RMT value drop indicates excitability increase.

**Figure 3. Longitudinal fMRI analysis in RGB style.** (**A**) 14 slices (Z = 15:10:145) of group RGB functional maps during left-hand active tasks for tDCS and sham groups. (**B**) The group RGB functional voxel statistics were calculated from individual fMRI maps within sham or tDCS groups. The detailed data statistics are presented in Supplementary Table 1.

**Figure 4. Granger causality graphs under left hand active task.** 22-node full brain G-causality graphs, for tDCS group (**A**) and sham (**B**) were calculated by MVGC engine. The changes of G-causality under rehabilitation are shown in 3-node sub-regions of motor cortex networks (right supplementary motor area, pre-motor and primary motor cortex) by post-training minus pre-training for real-tDCS (**C**) and sham (**D**) group. Red represents causality increase, blue represents decrease, thickness represents the magnitude, and * indicates $P < 0.05$. (**E-F**) show the causality changes in 5-node finger control networks (right Pre frontal cortex, pre-motor cortex, Insula, thalamus, and left cerebellum) for real and sham tDCS group, respectively. Note: left hand, left cerebellum and right motor cortex are stroke affected regions.

**Figure 5. Polarized GABA theory for tDCS effect.** External electric field *E* can re-orient polarized molecular GABA along field direction, that can either facilitate $GABA_A$ receptor activation (aligned with receptor pocket) or inhibit its activation (misaligned for most of the case), leading to $Cl^-$ inflow or gate shut, and final decrease or increase of neuron excitability, respectively.

# Reference

Adhikari MH, Deco G, Corbetta M. Reply: Defining a functional network homeostasis after stroke: EEG-based approach is complementary to functional MRI. Brain 2017; 140: e72–e72.

Agarwal SM, Shivakumar V, Bose A, Subramaniam A, Nawani H, Chhabra H, et al. Transcranial direct current stimulation in schizophrenia. Clin. Psychopharmacol. Neurosci. 2013; 11: 118–125.

Allman C, Amadi U, Winkler L, Wilkins AM, Filippini N, Kischka U, Stagg CJ J-BH. Ipsilesional anodal tDCS enhances the functional benefits of rehabilitation in patients after stroke. Sci. Transl. Med. 2016; 330re1.

Ang KK, Chin ZY, Wang C, Guan C, Zhang H. Filter bank common spatial pattern algorithm on BCI competition IV datasets 2a and 2b. Front. Neurosci. 2012; 6: 1–9.

Ang KK, Chin ZY, Zhang H, Guan C. Filter Bank Common Spatial Pattern (FBCSP) in brain-computer interface. Proc. Int. Jt. Conf. Neural Networks 2008: 2390–2397.

Ang KK, Guan C, Chua KSG, Ang BT, Kuah CWK, Wang C, et al. A large clinical study on the ability of stroke patients to use an EEG-based motor imagery brain-computer interface. Clin. EEG Neurosci. 2011; 42: 253–258.

Ang KK, Guan C, Phua KS, Wang C, Zhao L, Teo WP, et al. Facilitating effects of transcranial direct current stimulation on motor imagery brain-computer interface with robotic feedback for stroke rehabilitation. Arch. Phys. Med. Rehabil. 2015; 96: S79–S87.

Ang KK, Guan C, Phua KS, Wang C, Zhou L, Tang KY, et al. Brain-computer interface-based robotic end effector system for wrist and hand rehabilitation: results of a three-armed randomized controlled trial for chronic stroke. Front. Neuroeng. 2014; 7: 1–9.

Barnett L, Seth AK. The MVGC multivariate Granger causality toolbox: A new approach to Granger-causal inference. J. Neurosci. Methods 2014; 223: 50–68.

Bostan AC, Strick PL. The basal ganglia and the cerebellum: nodes in an integrated network. Nat. Rev. Neurosci. 2018; 19: 338-350.

Carter AR, Astafiev S V, Lang CE, Connor LT, Strube MJ, Pope DLW, et al. Resting Inter-hemispheric fMRI Connectiviyt Predicts Performance after Stroke. Ann. Neurol. 2010; 67: 365–375.

Cavaleiro P, Lomarev M, Hallett M. Modeling the current distribution during transcranial direct current stimulation. Clin. Neurophysiol. 2006; 117: 1623–1629.

Cooney J, Forte JD, Carter O. Neuropsychologia Evidence that transcranial direct current stimulation ( tDCS ) generates little-to-no reliable neurophysiologic effect beyond MEP amplitude modulation in healthy human subjects : A systematic review. Neuropsychologia 2015; 66: 213–236.

Van Delden ALEQ, Peper CLE, Beek PJ, Kwakkel G. Match and mismatch between objective and subjective improvements in upper limb function after stroke. Disabil. Rehabil. 2013; 35: 1961–1967.

Ebbesen CL, Brecht M. Motor cortex - To act or not to act?. Nat. Rev. Neurosci. 2017; 18: 694–705.

Elliott LT, Sharp K, Alfaro-Almagro F, Shi S, Miller KL, Douaud G, et al. Genome-wide association studies of brain imaging phenotypes in UK Biobank. Nature 2018; 562: 210–216.

Gladstone DJ, Danells CJ, Black SE. The fugl-meyer assessment of motor recovery after stroke a critical review of its measurement properties. - Neurorehabilitation and neural repair - 2002 - Gladstone, Danells, Black.pdf. Neurorehabil Neural Repair 2002a; 16: 232–240.

Gladstone DJ, Danells CJ, Black SE. The Fugl-Meyer Assessment of Motor Recovery after Stroke: A Critical Review of Its Measurement Properties. Am. Soc. Neurorehabilitation 2002b

Grefkes C, Fink GR. Reorganization of cerebral networks after stroke: New insights from neuroimaging with connectivity approaches. Brain 2011; 134: 1264–1276.

Hong X, Lu ZK, Teh I, Nasrallah FA, Teo WP, Ang KK, et al. Brain plasticity following MI-BCI training combined with tDCS in a randomized trial in chronic subcortical stroke subjects: A preliminary study. Sci. Rep. 2017; 7: 1–12.

Hummel FC, Steven B, Hoppe J, Heise K, Thomalla G, Cohen LG, et al. Deficient intracortical inhibition (SICI): During movement preparation after chronic stroke. Neurology 2009; 72: 1766–1772.

Jenkinson M, Beckmann CF, Behrens TEJ, Woolrich MW, Smith SM. Fsl. Neuroimage 2012; 62: 782–790.

Khedr EM, Ahmed MA, Fathy N, JC. R. Therapeutic trial of repetitive transcranial magnetic stimulation after acute ischemic stroke. Neurology. 2005; 65: 466–468.

Klamroth-Marganska V, Blanco J, Campen K, Curt A, Dietz V, Ettlin T, et al. Three-dimensional, task-specific robot therapy of the arm after stroke: A multicentre, parallel-group randomised trial. Lancet Neurol. 2014; 13: 159–166.

Langhorne P, Bernhardt J, Kwakkel G. Stroke rehabilitation. Lancet 2011; 377: 1693–1702.

Laverty D, Desai R, Uchański T, Masiulis S, Stec WJ, Malinauskas T, et al. Cryo-EM structure of the human α1β3γ2 GABAA receptor in a lipid bilayer. Nature 2019; 565: 516–520.

Lee JH Van Der, Beckerman H, Lankhorst GJ, Bouter LM. The responsiveness of the action research arm test and the Fugl-Meyer assessment scale in chronic stroke patients. J Rehab Med

2001; 33: 110–113.

Lindenberg R, Renga V, Zhu LL. Bihemispheric brain stimulation facilitates motor recovery in chronic stroke patients. Neurology 2010; 75: 2176-2184.

Masiulis S, Desai R, Uchański T, Serna Martin I, Laverty D, Karia D, et al. GABAA receptor signalling mechanisms revealed by structural pharmacology. Nature 2019; 565: 454–459.

Murase N, Duque J, Mazzocchio R, Cohen LG. Influence of Interhemispheric Interactions on Motor Function in Chronic Stroke. Ann. Neurol. 2004; 55: 400–409.

Murphy TH, Corbett D. Plasticity during stroke recovery: From synapse to behaviour. Nat. Rev. Neurosci. 2009; 10: 861–872.

Nachev P, Kennard C, Husain M. Functional role of the supplementary and pre-supplementary motor areas. Nat. Rev. Neurosci. 2008; 9: 856–869.

Nitsche MA, Cohen LG, Wassermann EM, Priori A, Lang N, Antal A, et al. Transcranial direct current stimulation : State of the art 2008. Brain Stimul. 2008; 1: 206–223.

Orlov ND, O'Daly O, Tracy DK, Daniju Y, Hodsoll J, Valdearenas L, et al. Stimulating thought: A functional MRI study of transcranial direct current stimulation in schizophrenia. Brain 2017; 140: 2490–2497.

Petoud S, Muller G, Moore EG, Xu J, Sokolnicki J, James P, et al. Symmetric Diffeomorphic Image Registration with Cross- Correlation: Evaluating Automated Labeling of Elderly and Neurodegenerative Brain. Med Image Anal 2008; 12: 26–41.

Pfurtscheller G, Lopes FH. Event-related EEG / MEG synchronization and desynchronization : basic principles. Clin. Neurophysiol. 1999; 110: 1842–1857.

Ramnani N. The primate cortico-cerebellar system: Anatomy and function. Nat. Rev. Neurosci. 2006; 7: 511–522.

Roelfsema PR, Holtmaat A. Control of synaptic plasticity in deep cortical networks. Nat. Rev. Neurosci. 2018; 19: 166–180.

Rossini PM, Burke D, Chen R, Cohen LG, Daskalakis Z, Iorio R Di, et al. Non-invasive electrical and magnetic stimulation of the brain , spinal cord , roots and peripheral nerves : Basic principles and procedures for routine clinical and research application. Clin. Neurophysiol. 2015; 126: 1071–1107.

Saleh A, Schroeter M, Jonkmanns C, Hartung HP, Mödder U, Jander S. In vivo MRI of brain inflammation in human ischaemic stroke. Brain 2004; 127: 1670–1677.

Seth AK. A MATLAB toolbox for Granger causal connectivity analysis. J. Neurosci. Methods 2010; 186: 262–273.

Smith SM. Fast robust automated brain extraction. Hum. Brain Mapp. 2002; 17: 143–155.

Smith SM, Nichols TE, Vidaurre D, Winkler AM, Behrens TEJ, Glasser MF, et al. A positive-negative mode of population covariation links brain connectivity, demographics and behavior. Nat. Neurosci. 2015; 18: 1565–1567.

Stagg CJ, Bachtiar V, O'Shea J, Allman C, Bosnell RA, Kischka U, et al. Cortical activation changes underlying stimulation-induced behavioural gains in chronic stroke. Brain 2012; 135: 276–284.

Thaut MH, McIntosh GC, R.R. Rice. Rhythmic facilitation of gait training in hemiparetic stroke rehabilitation. J. Neurol. Sci. 1997; 151: 207–212.

Tremblay S, Lepage JF, Latulipe-Loiselle A, Fregni F, Pascual-Leone A, Théoret H. The uncertain outcome of prefrontal tDCS. Brain Stimul. 2014; 7: 773–783.

Ward NS, Newton JM, Swayne OBC, Lee L, Thompson AJ, Greenwood RJ, et al. Motor system activation after subcortical stroke depends on corticospinal system integrity. Brain 2006; 129: 809–819.

Wong ROL, Ghosh A. Activity-dependent regulation of dendritic growth and patterning. Nat Rev Neurosci 2002; 3: 803–812.

Woodhead ZVJ, Kerry SJ, Aguilar OM, Ong Y-H, Hogan JS, Pappa K, et al. Randomized trial of iReadMore word reading training and brain stimulation in central alexia. Brain 2018; 141: 2127–2141.

Woolrich MW, Ripley BD, Brady M, Smith SM. Temporal autocorrelation in univariate linear modeling of FMRI data. Neuroimage 2001; 14: 1370–1386.

Wu J, Quinlan EB, Dodakian L, McKenzie A, Kathuria N, Zhou RJ, et al. Connectivity measures are robust biomarkers of cortical function and plasticity after stroke. Brain 2015; 138: 2359–2369.

Wu O, Christensen S, Hjort N, Dijkhuizen RM, Kucinski T, Fiehler J, et al. Characterizing physiological heterogeneity of infarction risk in acute human ischaemic stroke using MRI. Brain 2006; 129: 2384–2393.

Zhu S, Noviello CM, Teng J, Walsh RM, Kim JJ, Hibbs RE. Structure of a human synaptic GABAA receptor. Nature 2018; 559: 67-72.